# Understanding coil-to-globule transition of polymers with the aid of a novel cluster analysis technique.

T. E. Raptis

*Computational Applications Group, Division of Applied Technologies, National Centre for Science and Research "Demokritos", Athens, Greece*

Vasilios E. Raptis

*Molecular Systems Engineering Group, Centre for Process Engineering, Chemical Engineering Department, Imperial College London, London, England, UK*

Corresponding Author: Vasilios E. Raptis. E-mail: v.raptis@imperial.ac.uk

**Abstract**: In this article, a novel cluster analysis algorithm was employed in the study of polymer coil to globule transition via single chain Monte Carlo simulations. The algorithm, which has been recenlty applied in Molecular Dynamics simulations of atomistic systems that tend to phase separate [arXiv: 1307.7366 [cond-mat.soft]], provides us with a convenient means to map out the dynamics of "pearls" formation along the backbone chain together with extracting meaningful quantitative information about their shape and size distribution. Preliminary findings tend to favour a two-stage model of collapse kinetics, although a more complicated picture emerges when looking at the details of cluster formation along the chain.

## 1. Introduction

The drastic decrease in size of a polymer molecule when immersed in a bad solvent, the so called coil to globule transition, is a well-known phenomenon in polymer physics. This effect can be observed in a given solvent by changing the temperature, thus crossing different regions in the phase diagram. A change in the scaling behaviour is observed during the process: if the polymer chain dimensions obey a law of the form,

$\sim N^{v/2}$ where $N$ is the number of polymer segments, then a transition from $v = 6/5$ (good solvent), through $v = 1$ ($\Theta$-conditions) to $v = 2/3$ (bad solvent), takes place [1].

An understanding of the kinetics of the process and of the factors affecting it may be of great interest for advances and applications in a variety of fields, including nanotechnology and fabrication of nanocomposite materials, molecular biology etc. Pioneering work by de Gennes [2] suggested a two-stage pathway, of formation of polymer segment "blobs" ensued by their merger to minimise the surface energy between them and the solvent. The latter stage, though, can be further subdivided in a period where blobs stick together leading to a kind of a "sausage-like" formation and a period of further compactification due to rearrangements of the blobs.

Since then, the kinetics of coil-globule transition has been extensively studied at the levels of theory, experiment and simulations [3-21]. Some researchers tend to agree with de Gennes's initial suggestion about a two-stage model [3, 4, 6, 14, 15, 21], others claim the existence of a three- [7, 9, 17] or even four- [5, 11] stage mechanism encompassing processes of "pearl" coarsening or intermediate "molten globule" states, whereas still others doubt the existence of any distinct stages whatsoever [10, 12]. Experimentalists base their arguments on evidence for distinct relaxation times [3, 6, 10, 12, 14, 15] or different thermodynamically stable states on the way to collapse [9]. Theoreticians, either interpret the solution of mean-field based equations [7] or fit their models on a pre-assumed mechanism [11]. Molecular simulation studies [4, 5, 17, 21] benefit from the ability to compute global descriptors, as the radius of gyration or scaling exponents, or statistical local descriptors as the number of clusters or even to visualise directly the polymer conformations. It is worth noting, however, that the very definition of what is a "stage" of the process remains somewhat vague as for instance occasional density fluctuations of the initially expanded chain or slow rearrangements of the final globule are included in some of the above analyses, although one could argue that these do not constitute *sensu stricto* transitional phenomena.

In particular, Grosberg [3] used dynamic light scattering to observe two transitions of polystyrene in cyclohexane, from unknotted chain configurations to a globular and then on to a compact globule state in agreement with de Gennes's theory. Chu and Ying [6] on the other hand, studied the inverse process of polymer transition from the collapsed to the expanded chain, for similar systems, in an effort to explain the longer than theoretically predicted timescales observed, and concluded that chain knotting is not to be accounted for as a factor responsible for slow polymer-globule transitions. Wang et al [9] used laser light scattering to observe the size changes with temperature, of poly(N-isopropylacrylamide) collapsing and expanding in water and claimed the existence of four distinct polymer states, namely the ones of coil, crumpled coil, molten globule and globule.

Kayaman et al [10] studied via dynamic light scattering the collapsing process of poly(methyl methacrylate) in various solvents and observed a dependence of the involved time scales on molecular weight and temperature decrease in the quench experimental procedure. They claim that no evidence of distinct stages could be found and interpret previous results suggesting the opposite as due to intermolecular aggregation. Nakamura et al [12] studied similar systems using static light scattering and observed characteristic time scales of the process ranging from minutes to days, depending on the polymer molecular weight. On the other hand, Xu et al [14] observed a "crumpling" stage and a collapsing stage in the transition of poly(N-isopropylacrylamide), with timescales of just 12 ms and 270 ms respectively. Ye et al [15] also observed a two stage mechanism for various molecular weights of the same polymer and found that the first, fast, relaxation time is independent of the polymer chain length (about just 0.1 ms), an indication that it corresponds to local processes of cluster formation. The second, slow, relaxation time increases slightly with molecular weight (exceeding 0.6 ms) so that it should be attributed to global chain rearrangements.

At the level of theoretical models, Klushin [8] formulated a phenomenological theory of the "necklace" model in which a hierarchy of self-similar local relaxation processes is predicted, involving translation of clusters that coalesce to form larger aggregates until reaching the size and timescales pertaining to the polymer molecule as a whole. Apart from this interesting approach the author is careful to note that experiments are difficult to interpret as for instance inter-chain clustering may also take place, thus complicating the picture. In a similar direction, Halperin and Goldbart [11] proposed a phenomenological model that distinguished between a molecular weight independent stage of cluster formation, one of cluster growing at the expense of the "bridges" connecting them, causing the latter to stretch, and one of further growth of the clusters and elimination of the bridges, leading to the decrease of the chain dimensions.

Finally, at the level of simulations, Byrne et al [4] resorted to Langevin Molecular Dynamics simulations of Lennard-Jones chains comprising from 100 to 1024 units. To simulate the polymer collapse, the authors tweaked the attractive potential parameters to match values for the second virial coefficient that are theoretically known to correspond to a collapsed state. They claim that the collapse kinetics are affected by a variety of underlying mechanisms although two predominant effects can be generally observed, namely a rapid formation of small clusters and a slow cluster coarsening period. Kuznetsov et al [5] on the other hand, employed the Monte Carlo method applied on a lattice polymer model as a means to overcome the hurdle of long computational times and obtain reliable statistics for chain lengths of the order of 1000 segments. To trigger the polymer collapse, the authors tuned the interaction parameter, $\chi$, expressing the difference between polymer-solvent and polymer-polymer and solvent-solvent interaction energies, so as to emulate the bad solvent effect. The authors observe a molecular weight independent stage of clusters formation followed by a stage of clusters coarsening (instead of de Gennes's sausage model) and then a period of globule formation and compactification. In a subsequent theoretical analysis based on the numerical solution of the equations of the Gaussian self-consistent approach [7], the authors refer to the last two stages mentioned in [5] as one stage of globule optimisation and compactification, even one hard to distinguish from the preceding clustering coarsening phase – and propose a three-stage model.

Tri Thanh Pham et al [17] used the Brownian dynamics method and classified their observations in three stages, namely cluster formation, cluster coarsening and globule formation, which does not change qualitatively but can be accelerated due to the effect of hydrodynamic modes. Polson et al [18] carried out Monte Carlo simulations and used the energy mismatch between solvent and monomers to drive the transition process. Monte Carlo methods have also been employed to look at more complex systems as by Yuan et al [19] who studied how magnetic fields can induce the coil-globule transition in polyelectrolyte systems containing magnetic nanoparticles or to infer theoretical arguments about the nature and order of the polymer transition; the latter was also studied by Maffi et al [20] who extended the traditional discrete models to include vibrational modes. In this way they were able to show that entropic contributions can be incorporated in an analysis capable of reconciling conflicting theoretical models and experimental findings. Monte Carlo was also the method that Wang [21] used to study the effect of coil-globule transition to the polymer crystallisation; the author used a lattice model and decreased the temperature gradually to make the chain undergo a transition. According to the study, three stable states can be observed: coil, molten globule and globule.

The variety of the findings presented in the previous paragraphs probably reflects particular characteristics of the systems studied, experimental setup or assumptions underlying the theoretical explanations or employed simulation methods and models. To test the validity of these theories it is preferable to look at as high molecular weights as possible, so that theoretically predicted scaling laws apply. As the polymer collapse is a time-dependent process, Molecular Dynamics is the method of option. To overcome the hurdle of long time scales various techniques can be employed such as resorting to various levels of model coarse-graining. However, a convenient alternative exists, as some of the aforementioned studies suggest, namely employing Monte Carlo moves for a single polymer chain; this is especially true in the case of a chain *in vacuo* where the obstacle of surrounding molecules is removed. In such a simulation one can either change the temperature gradually to quench the chain or emulate a bad solvent environment by appropriately tuning the potential parameters. Another way to achieve the latter is by carefully selecting the range of intramolecular nonbonded interactions, to represent the solvent effect implicitly, as will be explained in the subsequent pages.

The advantage of this approach is that so-called "pivot" moves can be used to drastically alter the polymer's shape and size, so one can start with, say, an extended configuration and soon come up with a "crumpled" one and eventually, a collapsed chain, provided the intramolecular interactions are properly tuned to incorporate the bad solvent effect. The collected polymer samples constitute a kind of pseudo-temporal trajectory that can be useful in providing insights in the dynamics of coil to globule transition. Of course, it is not claimed that single chain MC can generally stand as a replacement for MD and its explicit time dependence or the ability to account for entanglement effects; it can, however, generate fastly a large set of configurations that represent the transient conformational subspaces visited by a polymer (even one of a realistically high molecular weight) on the way to its collapse, thus allowing a kind of broad overview of its trajectory.

The data sets thus obtained need to be processed so that quantitative information can be acquired, regarding the shape, size and temporal evolution of segment clusters formed along the chain. In principle, it should be trivial to locate the segment clusters and record their size. However, a certain arbitrariness enters in the "clusterness" criteria – as is the case in all problems involving the need to define clusters. This entails the risk of losing valuable information about phases of the process that simple visual inspection could reveal but would not suffice to treat it quantitatively.

In the next section, a new algorithm is presented, which can overcome the above arbitrariness by representing the examined data set as a geometrical object of lower dimensionality and defining the clusters on the basis of specific topological characteristics of that object. Then, the simulation method and model are presented, and, in the final Section, the results are discussed, with an emphasis on a particular illustration of the spatio-temporal distribution of the clusters; the latter is a straightforward output of our cluster identification technique and allows for deeper insights in the cluster formation and collapse mechanisms.

## 2. Methodology

### 2.1 Implementation of the cluster identification algorithm

A novel cluster analysis algorithm recently applied in simple Molecular Dynamics simulations of systems that tend to phase separate [22], addresses the issue of "fuzziness" in the cluster identification problem by using well defined characteristics of a one-dimensional geometrical object consisting of all the positions in question to define all possible clusters in it. The first step of the method is to define a short enough path (not necessarily the shortest) connecting all $N$ particles in question. One can then imagine a "walker" transversing the path and at each point calculating the size, e.g. the radius of gyration $R_g$, of a subset containing, say, the last $n = 50$ points encountered plus the next 50 to come, thus defining an "observation window" of size $w = 100$, sliding along the path as the walker proceeds. Where particles are clustered, the "window" size displays a local minimum, and adjacent maxima serve to separate it from other clusters. Additional criteria like a threshold cluster density, are particularly easy to incorporate so that clusters that are not closely enough packed can be discarded from the outset.

In general, an initial path that connects each particle position with the next one can be created e.g. by using a nearest-neighbors technique and then can be further refined. Ideally, one would search for an optimal solution to the Travelling Salesman Problem. Actually, this is not strictly necessary as it is sufficient if the path scans the volume occupied by a particular cluster before going to the next one. Therefore, the particular method does not need to find a minimal path but only filter out the ones that do not oscillate between neighbouring clusters. One can obtain such a path by resorting to simple well-known algorithms such as repeated exchanges of randomly selected point pairs until reaching a minimal length within a predetermined number of iterations.

In certain problems, especially those involving large data sets, more sophisticated algorithms may be needed to speed up the determination of a good enough path. In the particular case of single polymer chain computations, though, the polymer backbone provides a natural definition of the path, which renders the method's implementation even easier and quite fast. In this article, it is claimed that the specific method exhibits one more important advantage, namely it provides a very insightful way of visualising the dynamics of "pearls" formation by tracking the emergence and persistence with time, of the $R_g$ local minima along the polymer backbone. This advantage will be exploited in the subsequent paragraphs in order to understand the mechanisms underlying the coil to globule transition.

Regardless of the ways the path is formed, a cluster identification stage will follow: In it, the "walker" representing a serial scanning of the topologically one-dimensional object just created, computes the radius of gyration, $R_g$, over a set of nearest neighbour path points. In other words, a loop is performed over all points from $i = w$ to $N$ and the radius of gyration of the subset $\{i\text{-}w+1, i\text{-}w+2,\dots, i\}$ as a function of the objects' positions, $r_i$,

$$R_g(i) = \sqrt{\frac{\sum_{j=i-w}^{i}(r_i - r_{cm})^2}{w}}, \quad r_{cm} = \frac{\sum_{j=i-w}^{i} r_i}{w} \quad (1)$$

is tracked down with increasing $i$. The observation "window" width, $w$, is preselected by the user at input. The outcome is a set of values defining a $R_g$ curve over the path points. Location of successive minima immediately allows identification of each individual cluster. Neighbouring local maxima also define the edges of each cluster that is the path points at which a specific cluster "starts" or "ends". It should be stressed that the observation window, $w$, does not constitute a measure of the clusters' size. It is merely an auxiliary parameter that helps obtain a $R_g$ curve. A way to define an optimal window size is presented in Section 3.

At a subsequent post-processing stage, we have the freedom to apply a number of criteria that other existing cluster analysis methods cannot separate from the main body of the algorithm. In our simulations a further refinement was introduced consisting of a direct measurement of the pair of distances between each point of a given cluster's centre of mass and that of each neighbouring one. This allowed applying certain corrections corresponding to misclassification events such that all particle positions to be attributed to the most close cluster.

*2.2 Simulation details*

In this work, single molecule Monte Carlo simulations using the well-known MARTINI coarse-grained force-field [23] were carried out to study polyethylene chains of realistic molecular weight. To emulate the bad solvent effect, the following argument applied: An unperturbed polymer chain exhibits statistics quite similar to that of a "ghost chain" that can cross itself at segments that are situated far apart. We can exploit this fact to simulate unperturbed chains as chains in vacuum provided we know the minimum number of bonds two segments should be apart, in order to neglect their interaction (maximum range of local interactions).

To determine the local non-bonded interactions range, we have to think of the effect bonded and non-bonded forces have on the polymer size/shape and their possible competition. If the bonded geometry resembled an $sp^3$ tetrahedral one, then, bonded interactions alone tend to make the chain contract. A freely rotating chain, for instance, would exhibit a characteristic ratio equal to 2. If we allow for non-bonded interactions between segments separated by more and more bonds the excluded volume effect would start to be noticeable and help the chain expand. This way, at some intermediate values of local interaction range, $n$, the chain would resemble an unperturbed polymer, provided the total number of segments, $N$, would be much larger, $N \gg n$. This effect has served as a basis for single chain Monte Carlo simulations to validate new force-fields of silicon containing polymers [24].

On the contrary, if the geometry deviates largely from the tetrahedral structure and is closer to a linear one (as is the case with the MARTINI force-field), then bonded interactions alone have the opposite effect, making the chain expand. In the extreme case of a freely rotating chain with bond angles equal to 180º (as the MARTINI's equilibrium bond angles) the characteristic ratio would tend to infinity. This situation resembles a chain in a good solvent: its segments don't like each other and want to be surrounded by the solvent. As we allow more non-bonded pairs to contribute to the chain's energy, their attraction combined with entropic effects tends to win over the bonded interactions effect and bring the segments closer. At some point, the chain will resemble the unperturbed conditions and beyond that limit, it will tend to collapse, like a chain in a bad solvent where polymer segments prefer to be close to each other. From the above discussion it is implied that since we are using a model with a linear equilibrium angle we have to conduct trial simulations to determine the *lowest* possible maximum range of local interactions that would result in the polymer's collapse. To extend these interactions further to values comparable to the chain's degree of polymerisation would be both computationally uneconomical and wrong in principle as the chain would no longer obey the appropriate statistics.

In our simulations we modelled MARTINI-type polyethylene chains consisting of 3000 coarse-grained segments (corresponding to 9000 carbon atoms) subjected to different ranges of local interactions, expressed in number of segment-segment bonds, from $n = 4$ to up to 200 so that segments separated by $n+1$ or more bonds do not interact. Two kinds of pivot moves were employed with equal probability: internal rotations of randomly picked bonds to change the dihedral angles, and internal rotations about the axis normal to the plane defined by two randomly picked successive bonds to change the bond angles. Bond lengths were held fixed.

Chain dimensions are a function of temperature and of the solvent's nature. Following the argumentation presented in previous paragraphs we can emulate the bad solvent effect in two steps: we start with a temperature that is known to represent "theta conditions" for the modeled polymer (here: polyethylene) in a given solvent, and find the interactions range that would indeed reproduce the chain dimensions that are known to correspond to those conditions. Then, we can increase the local interactions range further to cause the collapse effect (provided the range remains much smaller than the number of polymer segments). Here, a temperature of 413 K representative of polyethylene's "theta conditions" in a variety of solvents [25], was chosen for all simulations. Each run started with an almost linear conformation and the chain was subjected to $10^6$ pivot moves.

### 3. Results and discussion

In agreement with the above reasoning, by gradually expanding the local interactions range we managed to obtain average chain dimensions corresponding to values of the characteristic ratio reported in the literature [25]. By increasing the interactions range further, a drastic decrease in the chain dimensions was observed. As an example, we take the case $n = 50$. The chain's radius of gyration with number of moves is shown in Fig. 1. Following a short equililbration period when the polymer "forgets" its initial stretched state and assumes realistic conformations, one can see a transition period of decreasing size and a final stabilisation period characterised by fluctuations around an average size, much smaller than the initial one. The transition period can be further divided into at least two stages on the basis of different rates of decrease in size. Interestingly, visual inspection of the generated configurations during the transition period showed the emergence of irregular ring-like clusters, Fig. 2, which would then collate further to form super-structures resulting in even more reduction in the chain's global dimensions. Whether these formations are "real" or merely an artefact due to the particular coarse-grained model will be resolved by future simulations using alternative representations of the polymer. To understand the clustering process leading to the chain collapse, the cluster analysis technique presented in Section 2.1 was employed. First, the "observation window" size, $w$, expressed in number of segments, had to be defined. To this purpose, cluster analysis of the whole simulation trajectory was carried out for different values of $w$ and the average cluster size was computed. That size increased with $w$ but at a slower rate with increasing $w$ until, for $w > 100$, a linear dependence set in, Fig, 3. We consider this as evidence that larger observation windows do not add new information, i.e. do not reveal larger clusters. We used values of $w = 100$ as well as 200 to cross-check results, in the rest of our analysis.

The number of clusters tended to increase with number of moves, whereas their size tended to decrease, Fig. 4. Indeed, the algorithm identifies as "clusters" any local minima of the observation window radius of gyration, so any occasional fluctuations in the segment density along the backbone may count. To avoid the "noise" of trivial segment density fluctuations, a density cutoff was applied. Once more compact and more stable clusters emerge the chain starts to shrink. Two broad categories of shrinking mechanisms can be suggested: a) cluster merger via translation along the chain (as an oscillation along a string would do) in opposite directions or simply because of growing larger; we sketchily call this, the "chemical" mechanism because it is mediated by the chain's chemical bonds; b) chain folding causing clusters that are topologically far apart to get closer in space and remain connected thanks to van der Waals or other non-bonded interactions. This kind of a tentatively called "geometrical" mechanism has been excluded by the very design of the present simulations based on the "ghost chain" concept, and will be studied in a forthcoming work of ours.

By isolating the "chemical" mechanism not only can we study its effect separately, but we can carry out the particular method of cluster analysis without having to redefine the path connecting the segments, as the general case would be [22]. However, such "geometrical" interactions are still present, albeit within a local range, as previously explained. These local interactions can help neighbouring clusters remain closer and form persistent structures. Every such event results in a drastic decrease in dimensionality of the remaining accessible configurational space, probably allowing for faster formation of new clusters.

To validate the above scenario, we plot all recorded cluster locations, expressed as the numbering index of every segment where a subset of monomers enclosed by the sliding "observation window" displays a minimum, versus the number of Monte Carlo moves. This is shown in Fig. 5 – only segments between the 1000-th and 2000-th one are represented on the vertical axis for the sake of clarity. Initially, clusters appear and disappear randomly at a fast rate. At about $10^5$ moves, four stable neighbouring clusters appear (segments between 1000 and 1200), another group of neigbouring clusters emerges in the (1800, 2000) segment zone after $10^5$ more moves and then, many more are formed, which make their way to the end of the simulation. Interestingly, the emergence of a new stable cluster eliminates the noisy pattern around it, a fact that we interpret as the result of more neighbouring segments "attracted" by it and contributing to its enlargement. Occasionally, other patterns can be seen, as *translations* of clusters along the chain, *splits* or *mergers*. Apparently, stabilisation of clusters takes place through gradual incorporation of nearby segments and/or fusion events that take place at a rate exceeding the one of fissions.

The above visualisation is straightforward thanks to the particular cluster analysis technique as it is literally a built-in characteristic thereof; use of other methods would render such a task more tedious. We expect this advantage to pave new paths of insightful investigations of polymer dynamics. Although we do not claim that the present results are able to provide a definite answer to the debate about a two- or three-stage kinetics, present indications seem to favour the first model when looking at the transition *per se* in terms of global descriptors of the polymer such as its radius of gyration, Fig. 1. Nevertheless, when looking at a local level, a more complex picture emerges with many overlapping stages marked by local transitions due to the frequent emergence of stable clusters until the remaining slack polymer is almost eliminated and the molecule enters a period of global stabilisation in its final shrunk state, Figs. 1 and 4. It seems that de Gennes' initial suggestion about blobs tending to stick together was close to reality, but this process takes place in a very complicated manner that could cause observable changes in relaxation times or in proxy measures of the system state and size, which vary depending on particular characteristics of studied systems, experimental setup etc., thus explaining conflicting evidence presented in the literature.

To completely elucidate the underlying mechanisms, further refinement would be needed with an emphasis, among others, on the "geometrical" mechanism. This can be done by redefining the range of local interaction in terms of a cutoff distance instead of number of bonds. Our future work will also focus on ways to extract more meaningful quantitative information from such data sets and on MC simulations with explicit solvents. Also, Molecular Dynamics simulations extended far beyond the 100 ns timescales, of polyethylene/alkane solutions near the LCST point are under way to verify the results herein reported.

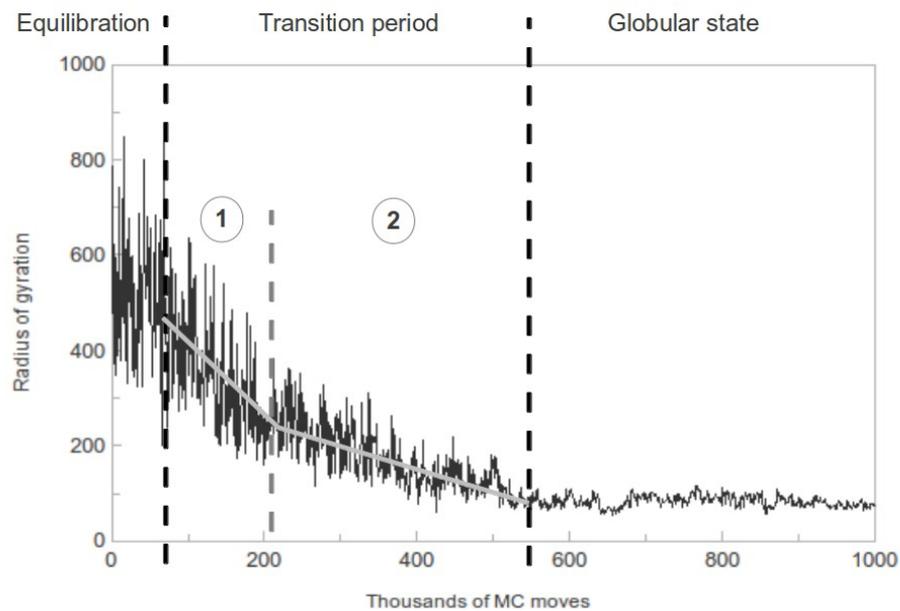

**Figure 1**. Radius of gyration of polymer chain, in Å, with number of MC moves. The vertical dashed lines delimit the period of transition to the globular state and the encircled numbers define two consecutive stages characterised by a different rate of decrease in polymer size, as shown by the superimposed grey lines.

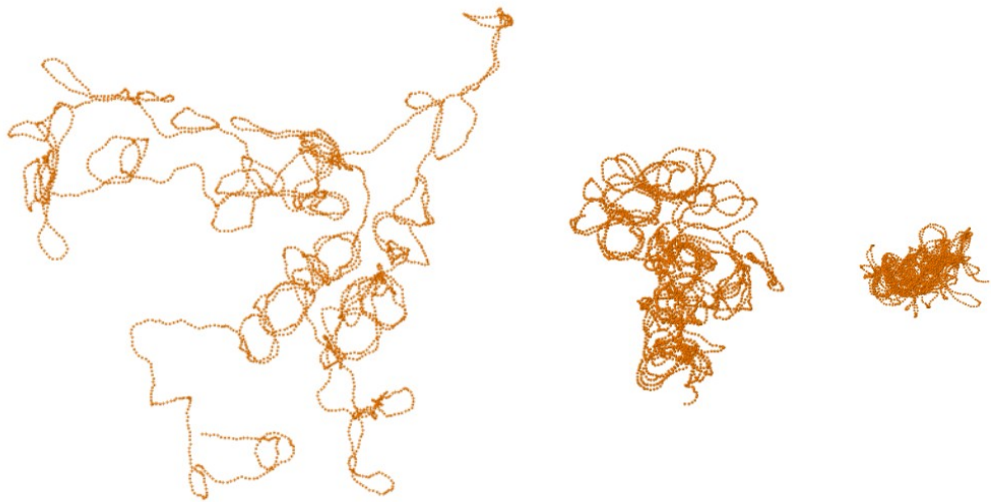

**Figure 2**. Three snapshots of the collapsing polymer chain; from left to right: 1.5 ✕ $10^5$, 5 ✕ $10^5$ and 9 ✕ $10^5$ MC moves.

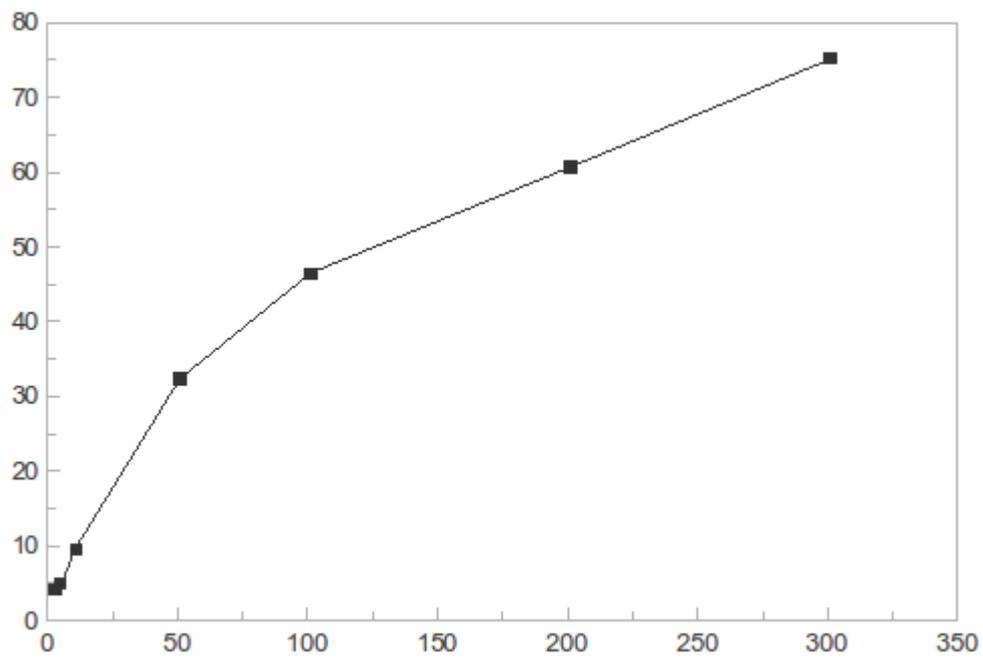

**Figure 3**. Dependence of average cluster size, in Å, on observation window size.

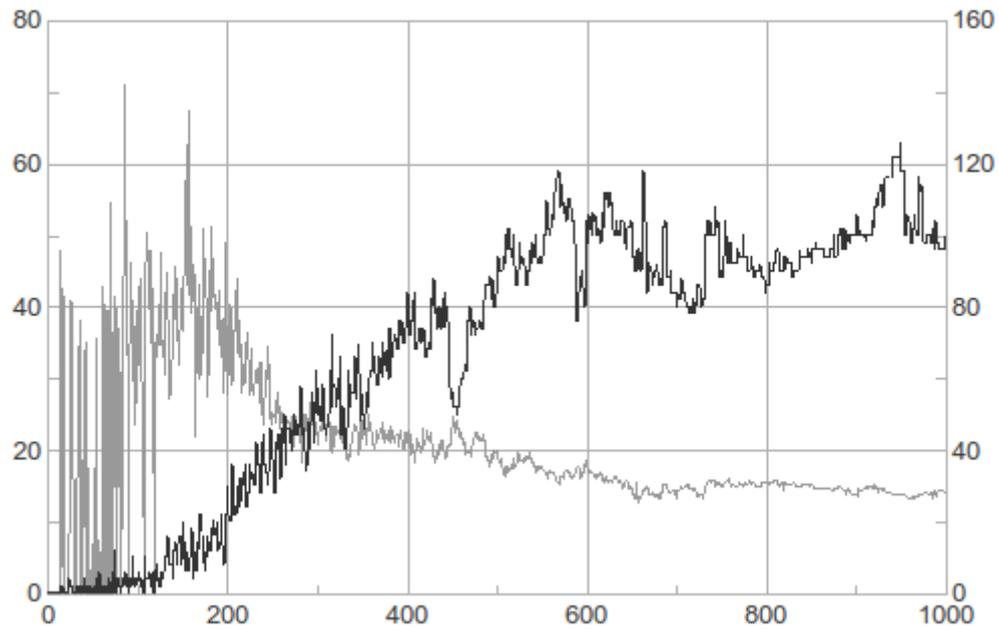

**Figure 4**. Number of clusters, dark grey line and left vertical axis, and average number of segments per cluster, light grey line and right vertical axis, with number of MC moves, using a density cutoff of $\rho = 3000$ relative to the total system density.

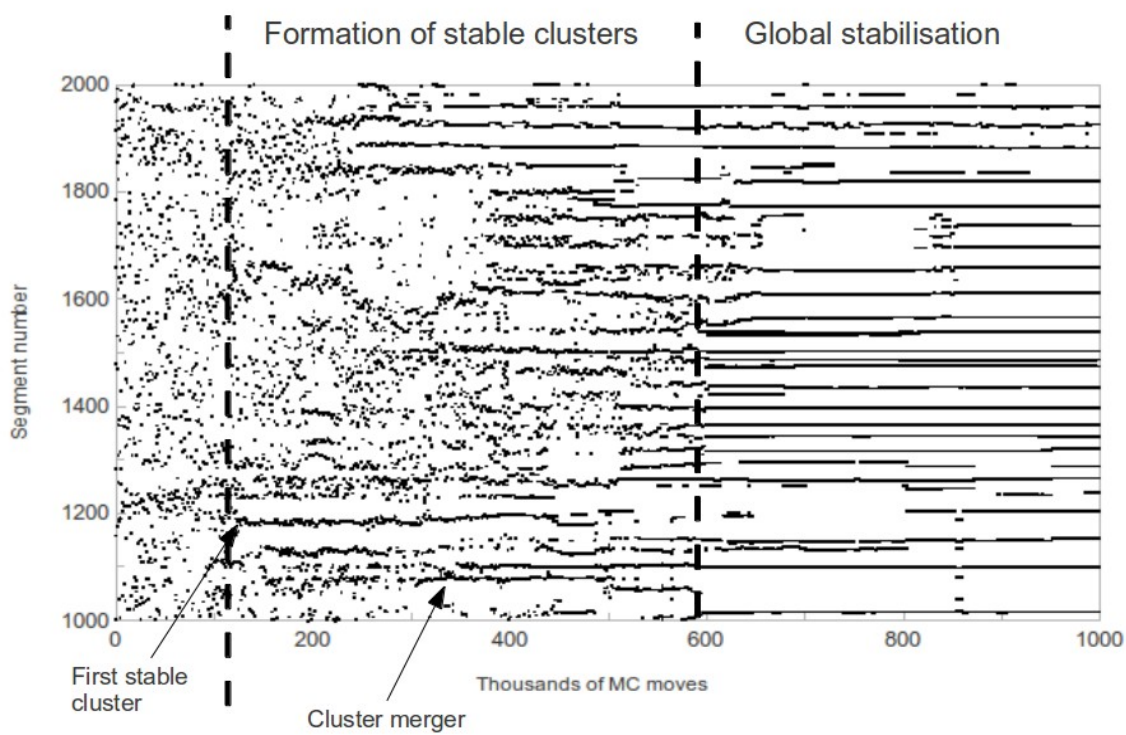

**Figure 5**. Location of emergent clusters along the polymer backbone as a function of MC moves. Horizontal axis: Thousands of moves. Vertical axis: restricted to segments from the 1000-th to the 2000-th for the sake of clarity. The vertical dashed lines delimit a transitional period marked by the stabilisation of locally shrunk conformations thanks to the emergence of new persistent clusters.